\documentstyle[letters]{mn}

\newif\ifAMStwofonts

\def\spose#1{\hbox to 0pt{#1\hss}}
\def\simlt{\mathrel{\spose{\lower 3pt\hbox{$\mathchar"218$}}
     \raise 2.0pt\hbox{$\mathchar"13C$}}}
\def\simgt{\mathrel{\spose{\lower 3pt\hbox{$\mathchar"218$}}
               \raise 2.0pt\hbox{$\mathchar"13E$}}}
\def\eg{{\rm e.g.}}
\def\ie{{\rm i.e.}}
\def\etal{{\rm et~al.}}
\def\vmi{\hbox{\it V--I\/}}

\ifoldfss
  \ifCUPmtlplainloaded \else
    \NewTextAlphabet{textbfit} {cmbxti10} {}
    \NewTextAlphabet{textbfss} {cmssbx10} {}
    \NewMathAlphabet{mathbfit} {cmbxti10} {} % for math mode
    \NewMathAlphabet{mathbfss} {cmssbx10} {} %  "   "    "
  \fi
  \ifAMStwofonts
    \ifCUPmtlplainloaded \else
      \NewSymbolFont{upmath} {eurm10}
      \NewSymbolFont{AMSa} {msam10}
      \NewMathSymbol{\upi}     {0}{upmath}{19}
      \NewMathSymbol{\umu}     {0}{upmath}{16}
      \NewMathSymbol{\upartial}{0}{upmath}{40}
      \NewMathSymbol{\leqslant}{3}{AMSa}{36}
      \NewMathSymbol{\geqslant}{3}{AMSa}{3E}

    \fi
  \fi
\fi % End of OFSS

\ifnfssone
  \newmathalphabet{\mathit}
  \addtoversion{normal}{\mathit}{cmr}{m}{it}
  \addtoversion{bold}{\mathit}{cmr}{bx}{it}
  \newmathalphabet{\mathbfit} % math mode version of \textbfit{..}
  \addtoversion{normal}{\mathbfit}{cmr}{bx}{it}
  \addtoversion{bold}{\mathbfit}{cmr}{bx}{it}
  \newmathalphabet{\mathbfss} % math mode version of \textbfss{..}
  \addtoversion{normal}{\mathbfss}{cmss}{bx}{n}
  \addtoversion{bold}{\mathbfss}{cmss}{bx}{n}
  \ifAMStwofonts
    \ifCUPmtlplainloaded \else
      \UseAMStwoboldmath
      \makeatletter
      \new@mathgroup\upmath@group
      \define@mathgroup\mv@normal\upmath@group{eur}{m}{n}
      \define@mathgroup\mv@bold\upmath@group{eur}{b}{n}
      \edef\UPM{\hexnumber\upmath@group}
      \new@mathgroup\amsa@group
      \define@mathgroup\mv@normal\amsa@group{msa}{m}{n}
      \define@mathgroup\mv@bold\amsa@group{msa}{m}{n}
      \edef\AMSa{\hexnumber\amsa@group}
      \makeatother
      \mathchardef\upi="0\UPM19
      \mathchardef\umu="0\UPM16
      \mathchardef\upartial="0\UPM40
      \mathchardef\leqslant="3\AMSa36
      \mathchardef\geqslant="3\AMSa3E
    \fi
  \fi
\fi % End of NFSS release 1

\ifnfsstwo
  \DeclareMathAlphabet{\mathbfit}{OT1}{cmr}{bx}{it}
  \SetMathAlphabet\mathbfit{bold}{OT1}{cmr}{bx}{it}
  \DeclareMathAlphabet{\mathbfss}{OT1}{cmss}{bx}{n}
  \SetMathAlphabet\mathbfss{bold}{OT1}{cmss}{bx}{n}
  \ifAMStwofonts
    \ifCUPmtlplainloaded \else
      \DeclareSymbolFont{UPM}{U}{eur}{m}{n}
      \SetSymbolFont{UPM}{bold}{U}{eur}{b}{n}
      \DeclareSymbolFont{AMSa}{U}{msa}{m}{n}
      \DeclareMathSymbol{\upi}{0}{UPM}{"19}
      \DeclareMathSymbol{\umu}{0}{UPM}{"16}
      \DeclareMathSymbol{\upartial}{0}{UPM}{"40}
      \DeclareMathSymbol{\leqslant}{3}{AMSa}{"36}
      \DeclareMathSymbol{\geqslant}{3}{AMSa}{"3E}
    \fi
  \fi
\fi % End of NFSS release 2

\ifCUPmtlplainloaded \else
  \ifAMStwofonts \else % If no AMS fonts
    \def\upi{\pi}
    \def\umu{\mu}
    \def\upartial{\partial}
  \fi
\fi

\title[The Distance to Supernova 1998aq in NGC$\,$3982]
{The Distance to Supernova 1998aq in NGC$\,$3982}
\author[Peter B. Stetson \& Brad K. Gibson]
       {Peter B. Stetson$^{1,3}$
       \& Brad K. Gibson$^2$ \\
        $^1$Dominion Astrophysical Observatory, 
       Herzberg Institute of Astrophysics,
	    National Research Council, 5071 W. Saanich Rd.,\\
	    $\;\,$Victoria, B.C., V9E~2E7, Canada \\
        $^2$Centre for Astrophysics \& Supercomputing, Swinburne University,
	    Mail \#31, P.O. Box 218, Hawthorn, Victoria, 3122, Australia\\
        $^3$Guest User, Canadian Astronomy Data Centre, which is operated by
	    the Dominion Astrophysical Observatory for the \\
	    $\;\,$Canadian National Research Council's Herzberg Institute of
	    Astrophysics
}
\date{Accepted 2001 ----- --.
      Received 2001 ----- --;
      in original form 2001 ----- --}

\pagerange{\pageref{firstpage}--\pageref{lastpage}}
\pubyear{2001}

\begin{document}

\maketitle

\label{firstpage}

\begin{abstract}
The distance to NGC$\,$3982, host galaxy to the Type~Ia supernova
SN$\,$1998aq, is
derived using 32 Cepheids discovered in archival multi-epoch
\it Hubble Space Telescope \rm observations.  Employing recent
Large Magellanic Cloud Cepheid period-luminosity relations and absolute
zero point, we find a distance to NGC$\,$3982 of 
20.5$\pm$0.8~($r$)$\pm$1.7~($s$)\,Mpc, including both random ($r$) and
systematic ($s$) uncertainties, and ignoring any metallicity dependency
in the Cepheid period-luminosity relation.
Still unpublished light curve photometry promises to make SN$\,$1998aq one
of
the most important calibrators for the Type~Ia supernova decline rate--peak
luminosity relationship.
\end{abstract}


\section{Introduction}

Corrected for the shape of their light curves, and calibrated via Cepheid
distances to their host galaxies, the peak luminosities of Type~Ia
supernovae (SNe) provide excellent standard candles for studying the
extragalactic distance scale.  Their extremely high peak luminosities
coupled with small intrinsic scatter lead to Type~Ia SNe being recognised
as perhaps the most powerful of available tools for deriving the Hubble
Constant (\eg, Phillips \etal\ 1999; Riess \etal\ 1999; Jha \etal\ 1999;
Gibson \etal\ 2000; Saha \etal\ 2001; Freedman \etal\ 2001; Gibson \&
Stetson 2001; Gibson \& Brook 2001).

Cepheid distances to ten Type~Ia SN-host galaxies are now available in the
literature --- NGC$\,$4414 (Turner \etal\ 1998), NGC$\,$3368 (Tanvir \etal\
1999),
NGC$\,$2841 (Garnavich et~al. 2001), 
and seven from the Saha \etal\ (2001, and references therein)
Type~Ia SN {\it Hubble Space Telescope (HST)} Calibration Project (IC~4182,
NGC$\,$5253, 4536, 4496A, 4639, 3627, and 4527).  Calibrating the corrected
peak luminosities of the respective SNe resident in each of these galaxies, 
and employing the Cal\'an-Tololo and CfA Type~Ia SN Hubble Diagrams to
derive the Hubble Constant $H_0$, yields 
$H_0$=72$\pm$2~($r$)$\pm$7~($s$)\,km\,s$^{-1}$\,Mpc$^{-1}$ -- 
including both total random ($r$) and systematic ($s$) uncertainties -- 
e.g. Gibson \& Stetson (2001).  An eighth galaxy observed as part of the
Saha \etal\ program, NGC$\,$3982, is host to the prototypical Type~Ia
supernova SN$\,$1998aq.  Of all the calibrators discussed to date,
SN$\,$1998aq
has perhaps the highest quality, in terms of multiwavelength
photometric precision and light-curve coverage (see Table~1 of
Gibson \etal\ 2000).

SN$\,$1998aq was discovered on 1998 April 13, by Mark Armstrong (Hurst
et~al.  1998), as part of the U.K. Nova/Supernova Patrol.  The host
galaxy, NGC$\,$3982, is a Seyfert-2 SAB(r)b: spiral, and a
possible\footnote{Its membership has been questioned by some (\eg, Tully
et~al. 1996).} member of the Ursa Major Cluster of galaxies.  Armstrong's
discovery took place approximately two weeks before the supernova reached
its $B$-band maximum.  Exquisite multi-colour photometry spanning the rise
and fall of the supernova light curve was subsequently collected by at
least one team, but at the time of submission of this manuscript the data
had not yet been made public.  As a result, the incorporation of
SN$\,$1998aq
into the Type~Ia SN calibration zero point is necessarily reserved for a
future paper.  Until then, we make available to the community our new
Cepheid-based distance determination to the host galaxy, NGC$\,$3982.

\section{Analysis}

NGC$\,$3982 was observed with the Wide Field Planetary Camera 2 (WFPC2)
at 17 epochs over the 53-day window spanning 2000
March 20 -- May 12.  Cosmic-ray split 2500\,s exposures in F555W (V; 12
epochs) and F814W (I; 5 epochs) were taken.
As noted earlier, this
galaxy is the eighth to be observed as part of the
Saha \etal\ Type~Ia SN Calibration Project ({\it HST} PID 8100;
Saha et~al. 2001, and references therein).

Data processing followed the precepts outlined in Gibson \& Stetson (2001,
and references therein).  Instrumental photometry was provided by ALLFRAME
(Stetson 1994), while a version of TRIAL (Stetson 1996) --- customized for
WFPC2 data and the fitting of Cepheid light curves in the $V$ and $I$
photometric bands --- was used for calibration and variable finding.  A
slightly refined version of the Stetson (1998) WFPC2 photometric zero
points and charge transfer corrections were employed, as was done in
Freedman \etal\ (2001), Gibson \& Stetson (2001), and Gibson \& Brook
(2001).  Our quoted systematic error budget includes a $\pm$0.07\,mag
component that allows for the uncertainties associated with our current
understanding of the spatial and temporal properties of the WFPC2 
charge-transfer inefficiencies.

Candidate variables returned by TRIAL were culled to a final set of 32
high-quality Cepheids.  The criteria employed in this selection were as
follows:  \begin{itemize} \item modified Welch \& Stetson (1993) index $>$
0.60 \item variable has data from a minimum of ten frame-pairs \item mean
$V$-magnitude $\left<V\right>$ fainter than 24.4$\,$mag \item mean colour
$0.40<
(\left<V\right>-\left<I\right>)<1.40$ \item period $>\,$10\,days 
\item semi-amplitude of fundamental
harmonic $>$0.22$\,$mag \item visual inspection of image \item visual
inspection of lightcurve \end{itemize} A selection on the ALLFRAME
image-quality index $\chi$ was considered, but the stars meeting the above
eight criteria were found to be completely normal in their $\chi$ values.
The properties of the 32 Cepheids passing this selection are listed in
Table~1.  Epoch-by-epoch photometry for each of the Cepheids, local
calibration standards, and accompanying light curves have been made
available at our {\it HST} Cepheid archive.\footnote{See \tt
http://astronomy.swin.edu.au/bgibson/H0kp\rm.}

\begin{table*}
\centering
\begin{minipage}{90mm}
\caption{Properties of Cepheids detected in NGC~3982.}
\begin{tabular}{@{}ccrrcccc@{}}
 ID & Chip & X$^a$ & Y$^a$ &  $<$V$>$    &     $<$I$>$    &        P       &
      $\mu_0$ \\
C01 & 1 & 240.3 & 245.6 & 26.98$\pm$0.04 & 25.98$\pm$0.07 & 16.77$\pm$0.21 &
	31.15 \\
C02 & 1 & 674.2 &  89.2 & 26.76$\pm$0.04 & 25.87$\pm$0.07 & 21.33$\pm$0.24 &
	31.54 \\
C03 & 2 & 125.2 & 319.8 & 26.08$\pm$0.03 & 25.33$\pm$0.06 & 37.05$\pm$1.59 &
	32.00 \\
C04 & 2 & 153.1 & 314.1 & 25.73$\pm$0.03 & 24.71$\pm$0.04 & 53.87$\pm$2.85 &
	31.50 \\
C05 & 2 & 177.9 & 496.6 & 26.53$\pm$0.05 & 25.70$\pm$0.08 & 25.83$\pm$0.59 &
	31.74 \\
C06 & 2 & 197.7 & 467.2 & 26.27$\pm$0.03 & 25.44$\pm$0.03 & 45.44$\pm$2.73 &
	32.28 \\
C07 & 2 & 220.0 & 742.7 & 26.49$\pm$0.04 & 25.57$\pm$0.06 & 32.02$\pm$1.03 &
	31.78 \\
C08 & 2 & 354.8 & 484.0 & 26.96$\pm$0.06 & 26.26$\pm$0.06 & 18.06$\pm$0.27 &
	31.97 \\
C09 & 2 & 356.4 & 691.9 & 26.64$\pm$0.03 & 25.88$\pm$0.06 & 19.90$\pm$0.30 &
	31.64 \\
C10 & 2 & 441.4 & 411.3 & 26.63$\pm$0.04 & 25.78$\pm$0.05 & 41.00$\pm$1.71 &
	32.44 \\
C11 & 2 & 463.2 &  -4.2 & 26.59$\pm$0.04 & 25.91$\pm$0.07 & 17.68$\pm$0.67 &
	31.63 \\
C12 & 2 & 471.9 &  32.7 & 27.14$\pm$0.05 & 26.43$\pm$0.18 & 12.47$\pm$0.34 &
	31.60 \\
C13 & 2 & 503.7 &  33.8 & 26.70$\pm$0.03 & 25.94$\pm$0.05 & 12.81$\pm$0.19 &
	31.07 \\
C14 & 2 & 504.4 & 287.4 & 26.67$\pm$0.05 & 25.54$\pm$0.06 & 29.08$\pm$1.84 &
	31.28 \\
C15 & 2 & 551.4 & -25.1 & 26.91$\pm$0.03 & 25.93$\pm$0.05 & 24.94$\pm$1.16 &
	31.70 \\
C16 & 2 & 578.5 & 390.6 & 26.73$\pm$0.03 & 25.55$\pm$0.05 & 38.12$\pm$0.59 &
	31.62 \\
C17 & 2 & 613.2 & 392.8 & 26.85$\pm$0.04 & 26.26$\pm$0.07 & 13.41$\pm$0.22 &
	31.72 \\
C18 & 3 & 245.7 & 504.8 & 26.04$\pm$0.04 & 25.07$\pm$0.05 & 31.60$\pm$1.25 &
	31.18 \\
C19 & 3 & 446.1 & 463.9 & 25.79$\pm$0.02 & 25.12$\pm$0.04 & 27.23$\pm$0.88 &
	31.47 \\
C20 & 3 & 446.4 & 591.2 & 26.10$\pm$0.04 & 25.51$\pm$0.07 & 25.01$\pm$1.38 &
	31.85 \\
C21 & 3 & 701.3 &  -1.6 & 26.80$\pm$0.05 & 25.68$\pm$0.06 & 30.04$\pm$2.92 &
	31.50 \\
C22 & 4 & 137.7 & 145.1 & 26.10$\pm$0.04 & 25.00$\pm$0.04 & 45.18$\pm$2.75 &
	31.42 \\
C23 & 4 & 192.5 & 200.0 & 26.61$\pm$0.04 & 25.61$\pm$0.06 & 21.61$\pm$0.68 &
	31.15 \\
C24 & 4 & 215.8 & 402.2 & 26.44$\pm$0.03 & 25.30$\pm$0.04 & 37.86$\pm$1.11 &
	31.42 \\
C25 & 4 & 227.3 & 316.8 & 26.77$\pm$0.04 & 25.75$\pm$0.05 & 17.66$\pm$0.39 &
	30.95 \\
C26 & 4 & 278.8 & 193.4 & 26.48$\pm$0.05 & 25.37$\pm$0.06 & 38.87$\pm$1.02 &
	31.57 \\
C27 & 4 & 297.2 & 272.5 & 26.65$\pm$0.03 & 25.81$\pm$0.06 & 21.09$\pm$0.50 &
	31.56 \\
C28 & 4 & 324.9 & 201.4 & 26.57$\pm$0.05 & 25.75$\pm$0.06 & 19.08$\pm$0.57 &
	31.38 \\
C29 & 4 & 417.8 & 277.9 & 26.43$\pm$0.05 & 25.17$\pm$0.05 & 40.89$\pm$1.18 &
	31.23 \\
C30 & 4 & 487.3 & 213.7 & 26.14$\pm$0.03 & 25.19$\pm$0.04 & 44.22$\pm$1.38 &
	31.79 \\
C31 & 4 & 505.6 &  14.8 & 26.43$\pm$0.04 & 25.32$\pm$0.06 & 45.41$\pm$3.94 &
	31.75 \\
C32 & 4 & 562.8 &  50.3 & 25.65$\pm$0.02 & 24.81$\pm$0.03 & 29.53$\pm$0.88 &
	31.02 \\
\end{tabular}
\vskip3.0mm
$^a$ Pixel coordinates are references with respect to {\it HST} archive
image {\tt u5ky0101r}.
\end{minipage}
\end{table*}

Figure~1 shows the apparent $V$- and $I$-band period-luminosity (PL)
relations for the 32 Cepheids in NGC$\,$3982 (upper and middle panels,
respectively).  Figure~2 shows the positions of these Cepheids
in the calibrated ($V$,\vmi) colour-magnitude diagam.
After Freedman \etal\ (2001), the apparent moduli of Figure~1, coupled
with the assumption of a standard reddening law, Large Magellanic Cloud
(LMC) absolute distance modulus $\mu_0$(LMC)=18.50$\pm$0.10\,mag, and the
new OGLE LMC apparent PL relations (Udalski \etal\ 1999), yield the
distribution of dereddened true moduli $\mu_0$ shown in the 
bottom panel of Figure~1.  The unweighted mean true modulus, based upon
all 32 Cepheids, is $<$$\mu_0$$>$=31.559$\pm$0.081~($r$), where the 
total random uncertainty ($r$) includes components due to photometry,
extinction, and dereddened PL fit, added in quadrature (corresponding
to $R_{\rm PL}$ in Table~7 of Gibson \etal\ 2000).\footnote{Enforcing 
a lower period cut of 25\,days reduces the Cepheid sample from
32 to 20, the unweighted mean true modulus for which is 
$<$$\mu_0$$>$=31.626$\pm$0.096~($r$), consistent with our 
favoured result of $<$$\mu_0$$>$=31.559$\pm$0.081~($r$).  This suggests
that PL bias of the magnitude encountered for {\it some} galaxies
(see \S~3.4 of Freedman \etal\ 2001) does not afflict our NGC$\,$3982
dataset, and so we employ the full set of 32 Cepheids in our
quoted final result.} It is reassuring
that the {\it median} of the distribution (31.594) agrees with the 
{\it mean} (31.559).  

\begin{figure}
\centering
\vspace{8.5cm}
\includegraphics{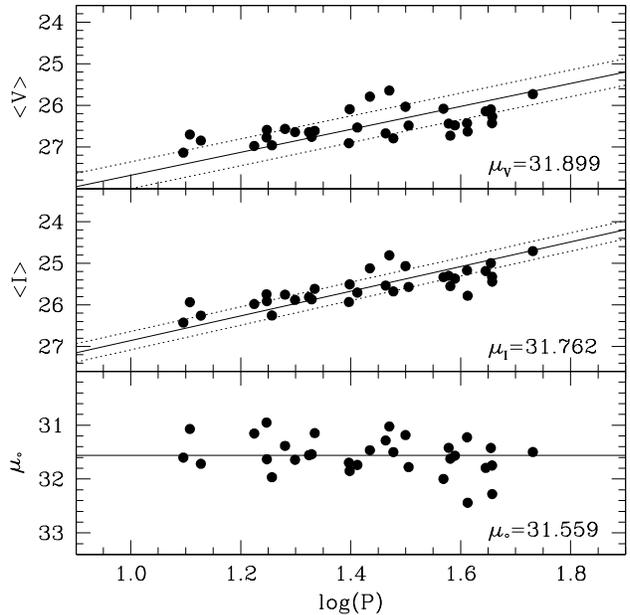}
\caption[]{Apparent period-luminosity relations in the $V$-
(\it upper panel\rm) and $I$-bands (\it middle panel\rm) based upon the 32
Cepheids discovered in NGC$\,$3982 (the properties of which are listed in
Table~1).  The solid lines are least-squares fits to this entire sample,
with the slope fixed to be that of the Udalski et~al. (1999) LMC PL
relations, while the dotted lines represent their corresponding $\pm
2\sigma$ dispersion - reflecting the width of the LMC instability strip.  
The inferred apparent distance moduli, ignoring
metallicity effects, are then $\mu_{\rm V}$=31.899$\pm$0.070 (internal) and
$\mu_{\rm I}$=31.762$\pm$0.056 (internal).  \it Lower Panel\rm:
Distribution of individually de-reddened Cepheid true moduli, as a function
of period.  The mean corresponds to $<$$\mu_\circ$$>$=31.559$\pm$0.081
(random).
\label{fig:fig1}}
\end{figure}

\begin{figure}
\centering
\vspace{8.5cm}
\includegraphics{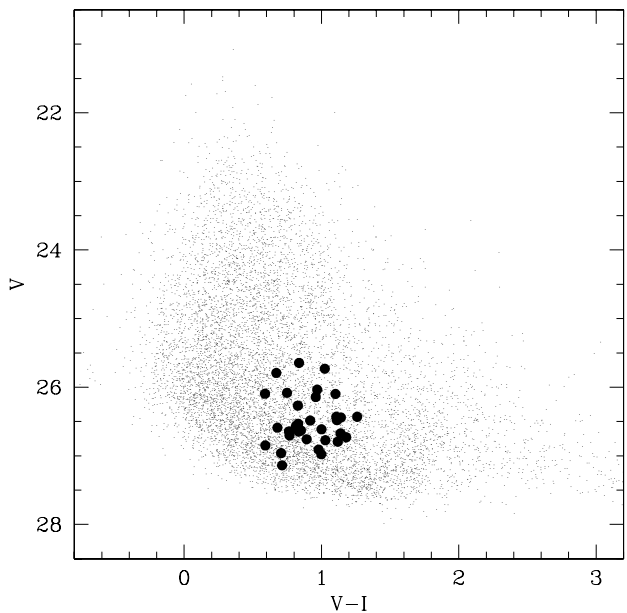}
\caption[]{Calibrated NGC$\,$3982 colour-magnitude diagram.  The filled
circles
correspond to the 32 Cepheids of Table~1.
\label{fig:fig2}}
\end{figure}

Our quoted result --- $<$$\mu_\circ$$>$=31.559$\pm$0.081\,($r$) ---
inherently assumes that the Cepheid PL relation is independent of
metallicity(\ie, $\gamma_{\rm VI}$=0.0\,mag\,dex$^{-1}$).  If one assumes
a metallicity dependence of the form $\gamma_{\rm
VI}$=$-$0.2\,mag\,dex$^{-1}$ (Freedman \etal\ 2001), though, a measure of
the mean metallicity for the WFPC2 Cepheid field is required to determine
the true metallicity-corrected distance modulus $\mu_Z$.  Unfortunately,
no H$\,${\small II} regions in this field have thus far been observed.
Adopting a conservative metallicity range of
$12+\log(\hbox{\rm O/H})=8.9\pm0.4$ (as we did for the analysis of the
NGC$\,$4527 dataset --- Gibson \& Stetson 2001) would increase the
inferred mean true modulus by $\sim$0.08\,mag.

The mean reddening inferred from the 32 Cepheids is
$E(\vmi)=0.138$$\pm$0.029\,mag (internal), of which only
\hbox{$\sim\,$0.019\,mag} is due to foreground Galactic extinction
(Schlegel \etal\ 1998).  While it might be tempting to assign a field mean
metallicity based upon an assumed {\it intrinsic\/} relationship between
$E(\vmi)$ and 12$+$$\log$(O/H), unfortunately --- as already noted in
Gibson \& Stetson (2001) --- for 0.1$\simlt$$E(\vmi)$$\simlt$0.3 (a regime
which the NGC$\,$3982 Cepheids inhabit), there is no trend seen in the
data, with metallicities 8.50$\simlt$12$+$$\log$(O/H)$\simlt$9.35 equally
likely.  The {\it true\/} intrinsic extinction in the vicinity of
SN$\,$1998aq (as opposed to assuming that the reddening appropriate to the
mean Cepheid necessarily applies to SN$\,$1998aq) will only become
apparent once the full multi-colour photometry of the supernova becomes
available.  

While its membership in the Ursa Major Cluster has not been established
unequivocally (Tully \etal\ 1996), it is perhaps telling that an independent
Tully-Fisher analysis leads to a mean cluster distance modulus of 31.48\,mag
(Freedman \etal\ 2001).  This latter value is consistent with our new
SN$\,$1998aq distance --- 31.559$\pm$0.081\,($r$)\,mag.  

Without access to calibrated multicolour photometry (or Cepheids), Vink\'o
\etal\ (1999) attempted to infer a distance to SN$\,$1998aq using the Riess
\etal\ (1999) `snapshot technique'.  Because of the lack of data available
at the time, deriving an accurate value of the reddening along the
line of sight was a near impossibility.  This uncertainty in the
extinction is responsible for an equal uncertainty in their SN$\,$1998aq
distance determination of 30.89$\pm$0.56\,mag.  Vink\'o \etal\ stress that
the true modulus is probably greater than their quoted value.  Regardless,
the original Vink\'o \etal\ value is consistent with our improved result,
at the $\sim$1.2\,$\sigma$ level.

We have not yet mentioned the systematic error budget.  After
Gibson \& Brook (2001; Table~2) and Freedman \etal\ (2001), 
seven sources of systematic uncertainty in the Hubble constant are
considered 
significant, including the LMC zero point, crowding, large scale
bulk flows, the metallicity dependence of the Cepheid PL relation, the WFPC2
zero points, Cepheid reddening values, and bias in the PL relation.  In
determining
the distance to an individual galaxy, all these error sources are equally
applicable
{\it except\/} the possibility of large-scale bulk flows.  In
quadrature,
the remaining systematic error budget ($s$)
amounts to 0.179\,mag.  In combination
with the already stated total random error budget, our final value for the 
distance to SN$\,$1998aq in NGC$\,$3982 is
$$
<\mu_\circ> = 31.559\pm 0.081\,(r)\pm 0.179\,(s)\,{\rm mag},
$$
assuming that a zero slope is the best available estimate for the
metallicity dependence of the Cepheid PL relation --- \ie,
$d$=20.5$\pm$0.8~($r$)$\pm$1.7~($s$)\,Mpc.  If the slope of the metallicity
dependence is in fact $-$0.2$\,$mag$\,$dex$^{-1}$, the modulus is increased
by $\sim$0.08\,mag for reasonable assumptions about the metallicity of
young stars in NGC$\,$3982.

\section{Summary}

A Cepheid-based distance to NGC$\,$3982, host to supernova SN$\,$1998aq, has
been derived using the same software pipeline and variable-finding algorithms 
employed throughout the {\it HST} Key Project on the Extragalactic 
Distance Scale.  Ignoring any potential metallicity dependence in
the Cepheid period-luminosity relation (expected to be on the order of
4\% for NGC$\,$3982), we determine a distance to the supernova (and galaxy)
of 20.5$\pm$0.8~($r$)$\pm$1.7~($s$)\,Mpc.  Our result is consistent with the
Tully-Fisher distance to the Ursa Major Cluster (19.8\,Mpc), suggesting
that NGC~3982 is indeed a cluster member.  Still unpublished 
multi-colour light curve photometry promises to make SN$\,$1998aq one of the
most important calibrators for the extragalactic distance scale.

\section*{Acknowledgments}

We would like to thank the staff of the Canadian Astronomy Data Centre for
providing an excellent service.

\bigskip
\noindent
{\it Note added in proof: } After submission, a preprint from Saha et~al.
(2001, ApJ, submitted, astro-ph/0107391) appeared which quotes an NGC~3982
distance modulus of $\mu_0$=31.72$\pm$0.14, based upon an independent 
DoPHOT and HSTphot analysis of the data described herein.  The distance
found by our ALLFRAME$+$TRIAL analysis - 31.56$\pm$0.08\,(r)$\pm$0.18\,(s) -
is tied to the Udalski et~al. (1999) LMC PL relations, while the Saha et~al.
result is based upon those of Madore \& Freedman (1991, PASP, 103, 933).
As discussed in Freedman et~al. (2001) 
however, the former are to be preferred over
the latter.  Regardless, if we were to adopt the Madore \& Freedman
relations, our inferred NGC~3982 distance modulus would become
31.73$\pm$0.08\,(r)$\pm$0.18\,(s), indistinguishable from that of
Saha et~al.

Despite these near-identical true distance moduli, there are differences
in the analyses - our apparent $V$- and $I$-band moduli are 0.13\,mag
and 0.07\,mag smaller than those of Saha et~al. - these differences 
are ameliorated by the inferred reddening $E(V-I)$, 
ours being 0.06\,mag lower
than that of Saha et~al ($E(V-I)$=0.05 versus 0.11,
for the same Madore \& Freedman 1981 LMC PL relations).  

A detailed comparison of the photometric differences between the two
analyses will have to wait until standard star photometry from the 
Saha et~al. group becomes available.  It should be noted though that such
Cepheid photometry differences are not uncommon (Gibson et~al. 2000;
Table~3).

\label{lastpage}

\end{document}